\documentclass[conference]{IEEEtran}
\IEEEoverridecommandlockouts
\usepackage[letterpaper, top=0.8in, bottom=1.04in, left=1in, right=1in]{geometry}
\usepackage{pgfplots}
\pgfplotsset{compat=1.18}
\usepackage{graphicx}
\usepackage{array}
\usepackage{cite}
\usepackage{amsmath,amssymb,amsfonts}
\usepackage{algorithmic}
\usepackage{textcomp}
\usepackage{xcolor}
\usepackage{float}
\usepackage{tikz}
\usetikzlibrary{arrows.meta, positioning, shapes.geometric}
\usepackage{amsmath}
\usepackage[table]{xcolor}
\usepackage{amssymb}
\usepackage{eso-pic}
\usepackage[hidelinks]{hyperref}
\begin{document}

\title{Towards Lightweight IoT Intrusion Detection: Feature Optimization and Explainable Machine Learning for Resource-Constrained Environments\\
}
\author{
\IEEEauthorblockN{Shaker Nawasra}
\IEEEauthorblockA{
School of Information Security \& Applied Computing \\
Eastern Michigan University \\
Ypsilanti, USA \\
snawasra@emich.edu
}
\and
\IEEEauthorblockN{Munther Abualkibash}
\IEEEauthorblockA{
School of Information Security \& Applied Computing \\
Eastern Michigan University \\
Ypsilanti, USA \\
mabualki@emich.edu
}
}
\maketitle

\begin{center}
\small
\textbf{DOI:}
\href{https://ieeexplore.ieee.org/abstract/document/11755189}
     {10.1109/Cyber-AI70573.2026.11755189}
\end{center}

\AddToShipoutPictureFG*{%
  \AtPageLowerLeft{%
    \put(\LenToUnit{1in},\LenToUnit{0.25in}){%
      \parbox[b]{\textwidth}{%
        \footnotesize
        \textcopyright\ 2026 IEEE. Personal use of this material is
        permitted. Permission from IEEE must be obtained for all
        other uses, in any current or future media, including
        reprinting/republishing this material for advertising or
        promotional purposes, creating new collective works, for
        resale or redistribution to servers or lists, or reuse of
        any copyrighted component of this work in other works.%
      }%
    }%
  }%
}

\begin{abstract}
Machine learning-based intrusion detection systems (IDSs) can improve IoT security but often rely on large feature sets that increase computational overhead and limit deployment on resource-constrained devices. This paper presents a resource-efficient and explainable feature optimization framework for lightweight IoT intrusion detection using the CICIoT2023 dataset. A Random Forest model was used to rank feature importance and identify compact feature subsets. Experimental results show that a three-feature model achieved an MCC of 0.9622 while reducing the feature space by 92.1\%, whereas a six feature model achieved an MCC of 0.9675 with an 84.2\% reduction. Furthermore, a fifteen-feature model achieved an MCC of 0.9744, preserving 99.97\% of the performance of the full 38-feature baseline model while reducing the feature space by 60.5\%. SHAP-based explainability analysis validated the selected features and provided interpretable insights into model decisions. The results demonstrate that lightweight, explainable, and high-performance IoT intrusion detection can be achieved using only a small subset of carefully selected features, significantly reducing resource requirements with minimal loss of detection effectiveness.
\end{abstract}
\vspace{0.5em}

\section{Introduction}
The increasing adoption of Internet of Things (IoT) devices has expanded the attack surface of modern networks, making intrusion detection systems (IDSs) essential for protecting IoT environments against evolving cyber threats\cite{s23135941}. Machine learning (ML)-based IDSs have demonstrated strong detection capabilities; however, many existing approaches rely on large feature sets and computationally intensive models that increase data collection, processing overhead, and deployment complexity\cite{sarhan2022featureanalysismachinelearningbased,Ma2025}. Such requirements limit their practicality in resource-constrained IoT environments with restricted processing power, memory, and energy resources \cite{benaddi2025lightweight,fi17110529}.

Feature optimization has emerged as an effective approach for reducing computational overhead while maintaining detection performance. Previous studies have shown that removing redundant and irrelevant features can improve efficiency and simplify model deployment\cite{Nazir2023FeatureSelection,s24165223}. In addition, explainable artificial intelligence (XAI) techniques such as SHAP can improve transparency by providing interpretable insights into model decisions\cite{Sharma2025}.

This paper presents a resource-efficient and explainable feature optimization framework for lightweight IoT intrusion detection using the CICIoT2023 dataset. A Random Forest model is employed to rank feature importance and identify compact feature subsets, while SHAP is used to explain prediction behavior. The experimental results demonstrate that a three-feature model achieves an MCC of 0.9622 while reducing feature-processing requirements by 92.1\%, and a fifteen-feature model preserves 99.97\% of the baseline performance while reducing feature-processing requirements by 60.5\%. These findings demonstrate that effective and explainable IoT intrusion detection can be achieved using only a small subset of carefully selected features, improving deployment feasibility in resource-constrained environments.

\section{Related Work and Research Gap}
Recent studies have explored feature selection and
machine learning techniques to improve IoT intrusion
detection performance. Ma et al. \cite{Ma2025}
combined feature selection with large language model
fine-tuning, while Alzahrani et al. \cite{alzahrani2024lightweight}
proposed a lightweight deep learning-based IDS for fog
computing environments. Chen et al. \cite{s24165223}
integrated explainable AI and feature selection to improve
model interpretability.

Although these approaches achieved high detection
accuracy, most studies focus primarily on classification
performance and often rely on accuracy as the main
evaluation metric. Furthermore, limited attention has
been given to resource-aware optimization, model size,
training cost, inference efficiency, and MCC-based
evaluation. In addition, few studies combine feature
reduction, explainability, and resource-efficiency analysis
within a single framework.

\section{Proposed Framework}

\subsection{Data Preparation and Balancing}
The CICIoT2023 dataset was used to evaluate the proposed intrusion detection framework. Preliminary analysis was conducted on multiple merged files (Merged01–Merged63) to assess dataset consistency and class distribution. The analysis revealed that all merged files exhibited nearly identical characteristics, containing approximately 97.6\% malicious traffic and only 2.4\% benign traffic. Such severe class imbalance can bias machine learning models toward the majority class, potentially leading to misleading performance results. For example, a classifier may achieve high accuracy by predominantly predicting the majority class while performing poorly on minority-class detection. To mitigate this issue, the dataset was cleaned and balanced prior to model training. Furthermore, Matthews Correlation Coefficient (MCC) was adopted as the primary evaluation metric because it incorporates all confusion matrix components (TP, TN, FP, and FN), providing a more reliable assessment of intrusion detection performance than accuracy alone. Consequently, Merged24 was selected as the representative dataset for all experiments.

The selected Merged24 dataset originally contained 743,201 network traffic records and 40 attributes. To improve data quality and prevent biased model behavior, duplicate records (177,377 samples) were removed, infinite values were replaced with missing values, and rows containing missing values were discarded. In addition, the attribute Number was removed to prevent information leakage, as it functions as a record identifier rather than a meaningful network traffic feature. After preprocessing, the dataset contained 565,812 records and 39 columns, including 38 traffic features and one class label.

To improve deployment practicality, the original multi-class attack categories were converted into a binary classification problem by grouping all attack types into a single Malicious class and retaining normal traffic as Benign. Despite preprocessing, the dataset remained highly imbalanced, containing 548,018 malicious records and 17,794 benign records. Such imbalance can bias machine learning models toward the majority class and produce misleading evaluation results. Therefore, a balanced dataset was created by retaining all benign samples and randomly selecting an equal number of malicious samples, resulting in 35,588 records evenly distributed between both classes. The resulting dataset was used for feature ranking, feature selection, model training, performance evaluation, and SHAP-based explainability analysis.

\subsection{Random Forest-Based Feature Ranking}
Random Forest was selected as the baseline learning algorithm because it consistently achieved the best overall intrusion detection performance in our previous comparative study involving Decision Tree, Naïve Bayes, Logistic Regression, and Support Vector Machine (SVM) classifiers. In addition to its strong classification capability and robustness against overfitting, Random Forest provides reliable feature importance estimates based on impurity reduction across decision trees. Feature importance scores were used to rank network traffic attributes from most to least informative, and the model was subsequently evaluated using progressively larger feature subsets. This process enabled the identification of compact feature sets that maintain high detection performance while reducing computational complexity.

\subsection{SHAP Explainability Module}
To improve model transparency and interpretability, SHAP (SHapley Additive exPlanations) was integrated into the proposed framework. SHAP was selected because it provides both global and local explanations, enabling comprehensive interpretation of feature contributions to individual predictions and overall model behavior. A SHAP summary analysis was performed on the optimized feature subset to identify the most influential traffic characteristics affecting intrusion detection decisions. By providing interpretable explanations for model predictions, SHAP helps security analysts validate alerts, investigate potential false positives, and increase confidence, robustness, and trust in the intrusion detection process.
\section{Experimental Setup}

\subsection{Evaluation Metrics}
The proposed framework was evaluated using both detection-performance and resource-efficiency metrics. Detection performance was assessed using Accuracy, Precision, Recall, F1-Score, and Matthews Correlation Coefficient (MCC). Although Accuracy is widely reported in intrusion detection research, it can be misleading on imbalanced datasets. Therefore, MCC was adopted as the primary evaluation metric because it incorporates all confusion matrix components (TP, TN, FP, and FN), providing a more reliable and balanced assessment of classification performance. Precision, Recall, and F1-Score were also reported to evaluate false alarm rates and attack detection capability.

To assess deployment suitability for resource-constrained environments, Training Time, Inference Time, and Model Size were measured. These metrics quantify the computational cost associated with different feature subsets and enable analysis of the trade-off between detection performance and resource efficiency.

\subsection{Experimental Environment}
All experiments were implemented in Python using the Scikit-learn library. Feature selection, model training, performance evaluation, and SHAP analysis were performed using stratified 5-fold cross-validation, with results averaged across all folds. The baseline Random Forest model described in Section III-B was used throughout all experiments.

\section{Results and Discussion}

\subsection{Baseline Detection Performance}

The baseline Random Forest model was trained using all 38 network traffic features. Table~\ref{tab:baseline} summarizes the performance achieved using the complete feature set.

\begin{table}[!h]
\centering
\caption{Baseline Detection Performance of the Full 38-Feature Model}
\label{tab:baseline}
\begin{tabular}{lll}
\hline
\textbf{Metric} & \textbf{Value} & \textbf{Purpose} \\
\hline
Accuracy  & 98.72\% & Overall correctness \\
Precision & 99.88\% & False alarm control \\
Recall    & 97.56\% & Attack detection rate \\
F1-Score  & 98.71\% & Precision--Recall balance \\
\rowcolor{gray!15}
\textbf{MCC} & \textbf{0.9747} & \textbf{Balanced evaluation} \\
\hline
\end{tabular}
\end{table}

As shown in Table~\ref{tab:baseline}, the baseline model achieved an accuracy of 98.72\%, precision of 99.88\%, recall of 97.56\%, F1-score of 98.71\%, and an MCC of 0.9747. The high MCC value indicates strong and balanced classification performance across both benign and malicious traffic classes. Overall, these results demonstrate the effectiveness of the baseline model and establish a reference point for evaluating the impact of feature reduction on intrusion detection performance.

\subsection{Feature Importance Analysis}

\begin{figure}[!h]
    \centering
    \includegraphics[width=\columnwidth]{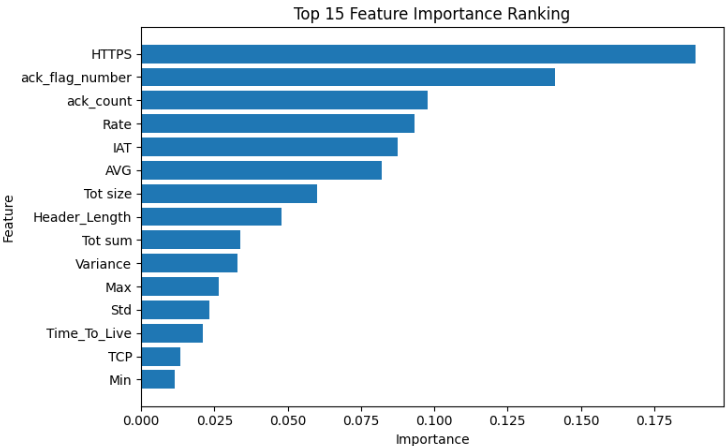}
    \caption{Feature importance ranking generated by the Random Forest model.}
    \label{fig:feature_importance}
\end{figure}

Figure~\ref{fig:feature_importance} 
presents the feature importance ranking generated by the Random Forest model. The results indicate that a small number of features contribute most of the predictive power. \textit{HTTPS}, \textit{ack\_flag\_number}, \textit{ack\_count}, \textit{Rate}, and \textit{IAT} were identified as the most influential features for distinguishing malicious traffic from benign traffic. This observation suggests that effective intrusion detection can be achieved without relying on the complete feature set.

\subsection{Performance versus Feature Count}

To identify the minimum feature subset required for effective intrusion detection, the model was evaluated incrementally using the top-ranked features, ranging from one feature to all 38 features. Figure~\ref{MCC vs Number of Selected Features} 

\begin{figure}[!ht]
    \centering
    \includegraphics[width=\columnwidth]{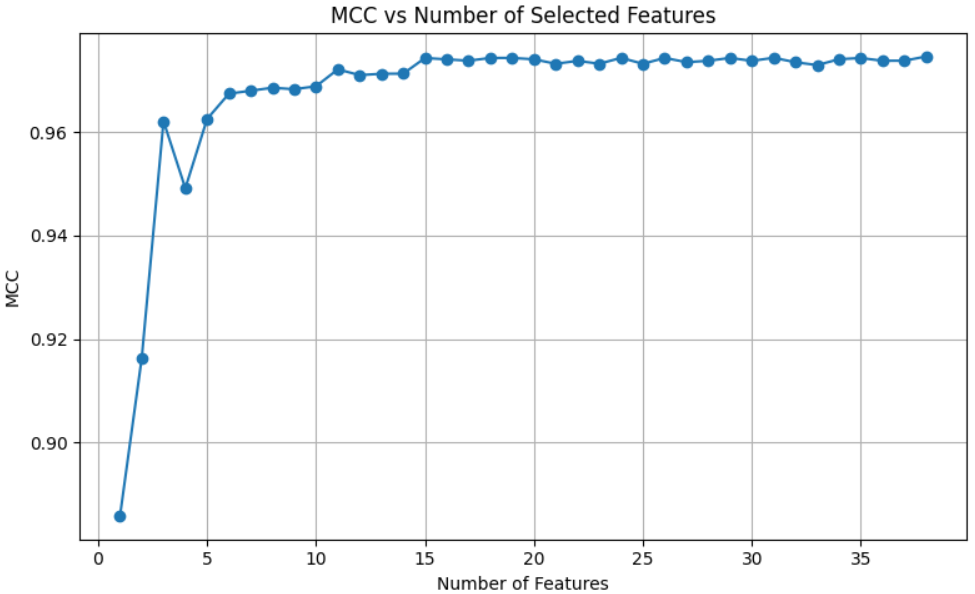}
    \caption{MCC vs Number of Features.}
    \label{fig:mcc-features}
\end{figure}

illustrates the relationship between the number of selected features and MCC performance. The results show a rapid performance improvement during the first few selected features, followed by gradual stabilization. Performance saturation was observed after approximately 15 features, where additional features produced only marginal improvements. Specifically, the 15-feature model achieved an MCC of 0.9744, compared to 0.9747 for the full 38-feature model.

\subsection{Resource-Aware Analysis}

Table II and Figure 3 illustrate the relationship between feature count, detection performance, and model complexity. As expected, increasing the number of selected features generally improved detection performance; however, the performance gains became marginal after approximately 15 features. The optimized 15-feature model achieved an MCC of 0.9744, which corresponds to approximately 99.97\% of the full 38-feature model performance (MCC = 0.9747), while reducing the feature space by 60.5\%.

Interestingly, Figure 3 shows that model size does not increase linearly with the number of selected features. Although additional features provide more information, the final model size is primarily determined by the structure and complexity of the decision trees generated during Random Forest training. An initial increase in model size was observed when highly informative features were introduced, resulting in deeper and more complex decision trees. However, as additional informative features became available, the model was able to construct more efficient decision boundaries, leading to a reduction in overall model size despite the larger feature set. Beyond approximately 15 features, the model size stabilized between 7 MB and 8 MB while detection performance remained nearly unchanged.

The results also demonstrate the feasibility of lightweight configurations. Using only the three most important features (HTTPS, ack\_flag\_number, and ack\_count), the proposed framework achieved an MCC of 0.9622 while reducing the feature space by 92.1\%. Similarly, the six-feature model achieved an MCC of 0.9675 with an 84.2\% feature reduction. These findings indicate that a significant portion of the original feature set can be eliminated while maintaining strong detection capability.

Overall, the results reveal a clear performance-resource trade-off and demonstrate that effective IoT intrusion detection can be achieved using substantially fewer features. This reduction decreases data collection requirements, feature processing overhead, storage complexity, and computational cost, making the proposed framework more suitable for deployment in resource-constrained IoT environments.

\begin{table*}[!h]
\centering
\caption{Resource-Aware Feature Selection Results}
\label{RESOURCE-AWARE FEATURE SELECTION RESULTS}
\resizebox{\textwidth}{!}{%
\begin{tabular}{cccccccccc}
\hline
Features & Accuracy & Precision & Recall & F1 & MCC &
Training (s) & Inference (s) & Latency (ms) & Model Size (KB) \\
\hline
3  & 0.9809 & 0.9945 & 0.9671 & 0.9806 & 0.9622 & 0.221 & 0.0378 & 0.0053 & 1461 \\
6  & 0.9836 & 0.9971 & 0.9699 & 0.9833 & 0.9675 & 0.522 & 0.0361 & 0.0051 & 9479 \\
11 & 0.9860 & 0.9983 & 0.9736 & 0.9858 & 0.9722 & 0.735 & 0.0368 & 0.0052 & 7556 \\
15 & 0.9871 & 0.9983 & 0.9758 & 0.9869 & 0.9744 & 0.627 & 0.0365 & 0.0051 & 7295 \\
38 & 0.9872 & 0.9988 & 0.9756 & 0.9871 & 0.9747 & 1.137 & 0.0531 & 0.0075 & 7532 \\
\hline
\end{tabular}}
\end{table*}

\begin{figure}[!h]
    \centering
\includegraphics[width=\columnwidth]{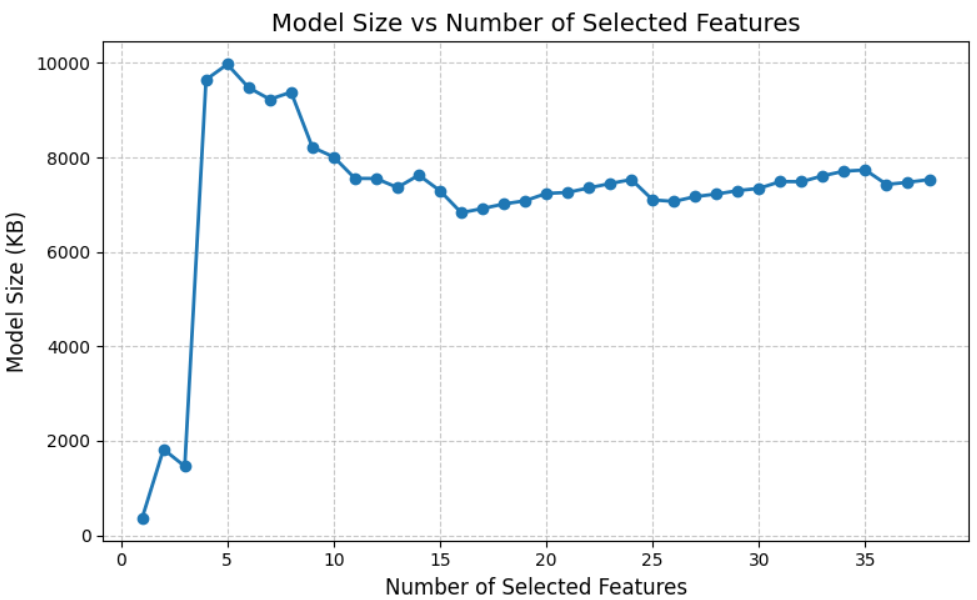}
    \caption{Model Size vs Number of Selected Features}
    \label{Model Size vs Number of Selected Features}
\end{figure}

\subsection{Explainability Analysis}

To improve model transparency and better understand the factors driving intrusion detection decisions, SHAP (SHapley Additive exPlanations) analysis was applied to the optimized 15-feature model. Figure 4 presents the SHAP summary plot, where each point represents an individual network traffic record. The horizontal position indicates the feature's contribution to the model prediction, while the color represents the feature value, ranging from low (blue) to high (red). Positive SHAP values increase the likelihood of classifying traffic as malicious, whereas negative SHAP values push the prediction toward benign traffic.

The results show strong agreement with the Random Forest feature ranking. The most influential features include ack\_flag\_number, HTTPS, IAT, Rate, and ack\_count, confirming that a small subset of features contributes most of the predictive power of the intrusion detection model.
Among these features, ack\_count exhibits the strongest positive impact on malicious predictions. Higher values of ack\_count (red points) are concentrated on the positive side of the SHAP scale, indicating that unusually large numbers of acknowledgment packets are strongly associated with malicious traffic. Similarly, higher values of Tot sum contribute positively to attack classification, suggesting that larger cumulative traffic volumes are frequently observed during malicious activities.

In contrast, lower values of HTTPS, Rate, AVG, and Header\_Length tend to increase the probability of malicious classification, while higher values of these features generally push predictions toward benign traffic. This behavior indicates that abnormal communication patterns, reduced traffic rates, or unusual packet characteristics are important indicators of suspicious network behavior within the CICIoT2023 dataset.

The SHAP analysis further reveals that the majority of prediction decisions are driven by a small number of highly influential features, while the remaining features contribute only marginally. This observation supports the feature optimization results presented in the previous sections and explains why substantial feature reduction can be achieved with minimal loss of detection performance. Specifically, the ability of the three-feature and six-feature models to maintain high MCC values is consistent with the SHAP findings, which show that predictive information is concentrated in a limited set of network traffic characteristics.

Overall, the SHAP results increase confidence in the model's decision-making process by providing interpretable explanations for intrusion detection outcomes. The analysis demonstrates that the proposed framework not only achieves high detection performance but also provides transparent and explainable security insights suitable for deployment in resource-constrained IoT environments.

\begin{figure}[!h]
    \centering
\includegraphics[width=\columnwidth]{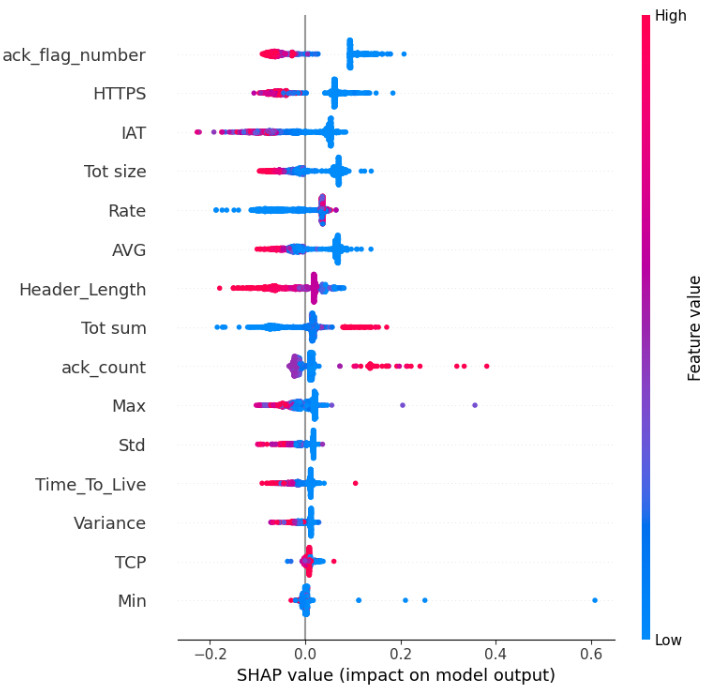}
    \caption{SHAP Summary Plot}
    \label{SHAP value (impact on model output}
\end{figure}

\subsection{Evaluation Beyond Accuracy: MCC-Based Comparison with Related Studies}
Accuracy is one of the most commonly reported evaluation metrics in intrusion detection research; however, it can be misleading when datasets are highly imbalanced. Consider a hypothetical dataset containing 1,000 network traffic records, where 990 records are benign and only 10 records are malicious. If a classifier predicts every record as benign, it correctly classifies 990 samples and incorrectly classifies 10 malicious samples. Although the model completely fails to detect any attack, it still achieves an accuracy of 99\%.

Table III presents this illustrative example. Despite the apparently excellent accuracy, the classifier fails to identify a single malicious record. This example demonstrates why accuracy alone is insufficient for evaluating intrusion detection systems. In cybersecurity applications, failing to detect malicious activity can have serious consequences despite high accuracy values. Therefore, this study emphasizes the Matthews Correlation Coefficient (MCC), which incorporates all elements of the confusion matrix (True Positives, True Negatives, False Positives, and False Negatives) and provides a more balanced assessment of classification performance. The use of MCC is particularly important for intrusion detection datasets, where class imbalance is common and accuracy may overestimate actual detection capability.

\begin{table}[!h]
\centering
\caption{Illustrative Example of Misleading Accuracy on an Imbalanced Dataset}
\label{tab:misleading_accuracy}
\begin{tabular}{lc}
\hline
\textbf{Metric} & \textbf{Value} \\
\hline
Total Records & 1,000 \\
Benign Records & 990 \\
Malicious Records & 10 \\
True Positives (TP) & 0 \\
True Negatives (TN) & 990 \\
False Positives (FP) & 0 \\
False Negatives (FN) & 10 \\
\rowcolor{yellow!25}
\textbf{Accuracy} & \textbf{99.0\%} \\
\rowcolor{red!20}
\textbf{MCC} & \textbf{0.0} \\
Recall & 0.0\% \\
\hline
\end{tabular}
\end{table}

For the scenario shown in Table III, the classifier fails to identify any malicious samples, resulting in an MCC value of 0 despite achieving an accuracy of 99\%.

\begin{equation}
\text{Accuracy}=\frac{990}{1000}=99\%
\end{equation}

The example highlights the importance of using MCC to assess intrusion detection performance. Motivated by this observation, Table IV compares recent IoT IDS studies based on MCC reporting, explainability(XAI), and feature reduction(FR). The comparison shows that most studies report Accuracy but not MCC, whereas the proposed framework integrates all three aspects within a single resource-aware IDS framework.

\begin{table}[!h]
\centering
\caption{Comparison of Recent IoT IDS Studies Based on Performance Evaluation and Optimization Criteria}
\label{tab:comparison}
\footnotesize
\renewcommand{\arraystretch}{1.0}

\begin{tabular}{p{2.0cm}cccc}
\hline
\textbf{Study} &
\textbf{Accuracy} &
\textbf{MCC} &
\textbf{XAI} &
\textbf{FR} \\
\hline

Alzahrani et al. (2024) \cite{alzahrani2024lightweight} & \checkmark & \cellcolor{red!20}\texttimes & \texttimes & \checkmark \\
Ma et al. (2025) \cite{Ma2025} & \checkmark & \cellcolor{red!20}\texttimes & \texttimes & \checkmark \\
Ullah et al. (2025) \cite{ullah2025comparative} & \checkmark & \cellcolor{red!20}\texttimes & \texttimes & \checkmark \\
Chen et al. (2024) \cite{s24165223} & \checkmark & \cellcolor{red!20}\texttimes & \checkmark & \checkmark \\
Benaddi et al. (2025) \cite{benaddi2025lightweight} & \checkmark & \cellcolor{red!20}\texttimes & \checkmark & \checkmark \\
Tayebi et al. (2025) \cite{tayebi2025rnnids} & \checkmark & \cellcolor{red!20}\texttimes & \texttimes & \texttimes \\
Akif et al. (2025) \cite{akif2025hybrid} & \checkmark & \cellcolor{red!20}\texttimes & \texttimes & \checkmark \\
Silva et al. (2024) \cite{silva2024iot} & \checkmark & \cellcolor{red!20}\texttimes & \texttimes & \checkmark \\
Susilo et al. (2025) \cite{susilo2025intelligent} & \checkmark & \cellcolor{red!20}\texttimes & \texttimes & \checkmark \\

\rowcolor{green!20}
\textbf{Proposed Framework} &
\textbf{\checkmark} &
\textbf{\checkmark} &
\textbf{\checkmark} &
\textbf{\checkmark} \\
\hline
\end{tabular}
\end{table}

\section{Conclusion}

This study presented a resource-aware and explainable feature optimization framework for lightweight IoT intrusion detection using the CICIoT2023 dataset. The experimental results demonstrate that effective intrusion detection can be maintained using only a small subset of the original network traffic features, significantly reducing computational requirements while preserving detection performance.

The findings reveal that a three-feature model, consisting of HTTPS, ack\_flag\_number, and ack\_count, achieved an MCC of 0.962 while reducing the original feature space by 92.1\%. Similarly, a six-feature model achieved an MCC of 0.967 with an 84\% feature reduction. These results indicate that the majority of the predictive information required for accurate intrusion detection is concentrated within a small number of highly informative features. Consequently, substantial reductions in data collection, storage, and processing requirements can be achieved with only a marginal decrease in detection effectiveness.

Furthermore, the optimized fifteen-feature model achieved an MCC of 0.9744 compared to 0.9747 for the full 38-feature baseline model, preserving approximately 99.97\% of the baseline performance while reducing the feature space by 60.5\%. The negligible difference between the reduced and full-feature models demonstrates that large feature sets are not necessary for high-quality intrusion detection. These findings highlight the feasibility of deploying machine learning-based IDS solutions on resource-constrained IoT devices, where memory, processing power, and energy availability are often limited.

In addition to maintaining strong detection performance, the proposed framework improves model transparency through SHAP-based explainability analysis. The explainability results validated the importance of the selected features and provided interpretable insights into the factors influencing intrusion detection decisions, increasing confidence in the reliability and trustworthiness of the proposed approach.

\section{Future Work}
Future work will focus on validating the proposed framework using additional IoT intrusion detection datasets to further assess model generalization and robustness. Real-world deployment on IoT and edge devices will also be investigated to evaluate practical performance in terms of latency, memory utilization, CPU consumption, and energy efficiency.

In addition, future studies may explore adaptive feature-selection techniques and compare lightweight machine learning models with deep learning approaches to better understand performance–resource trade-offs. The explainability component may also be extended through additional XAI methods and visualization tools to further improve transparency and support practical IoT security operations.

\bibliographystyle{IEEEtran}

\bibliography{references}
\end{document}